\catcode`\@=11

\newif\if@fewtab\@fewtabtrue
{\count255=\time\divide\count255 by 60
\xdef\hourmin{\number\count255}
\multiply\count255 by-60\advance\count255 by\time
\xdef\hourmin{\hourmin:\ifnum\count255<10 0\fi\the\count255}}
\def\ps@draft{\let\@mkboth\@gobbletwo
    \def\@oddhead{}
    \def\@oddfoot
       {\hbox to 7 cm{\tiny \versionno
       \hfil}\hskip -7cm\hfil\rm\thepage \hfil}
    \def\@evenhead{}\let\@evenfoot\@oddfoot}

\def\draftcite#1{\ifnum\draftcontrol=1#1\else{}\fi}

\def\@lbibitem[#1]#2{\item{}\hskip -3cm \hbox to 2cm
{\hfil$\scriptstyle\draftcite{#2}$}\hskip
1cm[\@biblabel{#1}]\if@filesw
     {\def\protect##1{\string ##1\space}\immediate
      \write\@auxout{\string\bibcite{#2}{#1}}}\fi\ignorespaces}

\def\@bibitem#1{\item\hskip -3cm \hbox to 2cm
{\hfil {\footnotesize\draftcite{#1}}}\hskip 1cm
\if@filesw \immediate\write\@auxout
       {\string\bibcite{#1}{\the\value{\@listctr}}}\fi\ignorespaces}

\global\def\draftcontrol{0}
\catcode`\@=12
\documentstyle[12pt,epsf]{article}

\newcommand{\bh}{black hole}
\newcommand{\cod}{co-ordinate}
\newcommand{\ee}{\end{equation}}
\newcommand{\eE}{{\rm e}}

\newcommand{\fourd}{four-dimensional}

\newcommand{\ii}{\rm i}
\newcommand{\ksz}{Kruskal-Szekeres}
\long\def\labl#1   {\label{#1}\ee\mbox{ }\\[-12 mm]\query{#1}\\[5 mm] }

\newcommand{\oned}{one-dimensional}
\long\def\query#1{\hskip 0pt{\vadjust{\everypar={}\small\vtop to 0pt{\hbox{}%
     \vskip -13pt\rlap{\hbox to 48.9pc{\hfil{\vtop{\hsize=8pc\tolerance=6000%
     \hfuzz=.5pc\rightskip=0pt plus 3em\noindent#1}}}}\vss}}}}%
\newcommand{\rd}{{\rm d}}
\newcommand{\st}{space-time}
\newcommand{\twod}{two-dimensional}

\newcommand{\zerod}{zero-dimensional}

\renewcommand{\labl}[1]{\label{#1}\end{equation}}

\begin{document}

\begin{flushright}  {~} \\[-26 mm] {\sf gr-qc/9407006} \\
{\sf NIKHEF-H/94-03} \\[1 mm]{\sf July 1994}
\end{flushright}

\begin{center}
\vskip 0.9cm

{\Large{\bf EXTENDED GEOMETRY }} \vskip 0.3cm
{\Large{\bf{OF BLACK HOLES}}}
\vskip 0.5cm

\vskip 0.7cm

{\large{ K.\ Peeters, C.\ Schweigert and }}  \vskip 0.2cm
{\large{ J.W.\ van Holten}} \vskip 0.2cm

NIKHEF/FOM

P.O.\ Box 41882

1009 DB Amsterdam NL


\end{center}
\date{ }

\vskip 1.6cm

\begin{quote}
{\bf Abstract.} \

We reconsider \st\ singularities in classical Einsteinian general
relativity: with the help of several new \cod\ systems we show that
the Schwarzschild solution can be extended beyond the curvature singularity at
$r=0$. The extension appears as an infinite covering of standard
Kruskal \st. While the \twod\ reduction of this infinite sequence of
\ksz\ domains obtained by suppressing the angular degrees of freedom
is still a topological manifold -- albeit one for which the metric structure
is singular on \oned\ submanifolds -- we obtain for the full \fourd\ geometry
the more general structure of a stratified variety.

\end{quote}

\vskip 1.3cm

In this letter we present some remarks on the structure of \st\ near
singularities in classical, pure Einsteinian general relativity, i.e.\ the
classical field theory for the metric structure defined by the Einstein
equations.
This theory in itself does certainly not describe all aspects of gravity,
especially not at small length scales where quantum corrections should be
taken into account.  However, for any quantum theory
of gravity a good understanding of the classical theory is helpful; in
addition, the theory has found important applications both in physics and in
mathematics.

In mathematical relativity \st\ is usually assumed (compare e.g.\ \cite{HE})
to be a smooth, connected, inextendable Hausdorff manifold $M$ with a smooth
Lorentz metric. (In fact, for
geodesics to be uniquely defined, it is only necessary to assume that the
metric is $C^{2-}$, i.e.\ differentiable with locally Lipschitz continuous
derivatives.) One definition of inextendability is to require that $M$ cannot
be embedded isometrically in a larger Lorentz manifold of the same dimension.
There is also the stronger notion of local inextendability \cite{HE}:
$M$ is called locally inextendable if there is no open set ${\cal U}$ in $M$
with non-compact closure that has an extension in which the closure of the
image of ${\cal U}$ is compact.

Inextendability is an important requirement coming from physics: an essential
aspect of general relativity is that the geometry of \st\ is subject
to dynamics. But then it is {\em compulsory} to consider an
inextendable \st\ manifold: taking an extendable \st\ amounts to stopping
dynamics at an arbitrary point. Such arbitrariness affects the validity of
any global result.

The geometry of \st\ is usually studied by means of geodesics. {}From the
point of view of physics this involves the idealization of `test particles'
which certainly is debatable. {}From a mathematical point of view geodesics are
a rather natural tool since they are known to encode most of the
geometric information about $M$. Therefore, in this letter, completeness of
(time-like) geodesics will be taken as the criterion for inextendability.

The notion of inextendability depends, of course, strongly on the
other requirements one imposes on $M$: weakening one of them may imply
that a manifold that is inextendable in the original axiomatic setting becomes
extendable in the new one.
Below we present some arguments that requiring $M$ to be a
smooth manifold with a smooth metric is too restrictive and that one has to
include singularities in the description of \st. This in turn will allow for
the extension of Schwarzschild \st\ we are going to present.

The main mathematical observation behind our results is the simple fact that
the solution of a differential equation -- in our case the geodesic equation --
can be regular even if the coefficients of this equation, given by the metric
and its derivatives, have singular points.

Proposals to extend the Schwarzschild solution beyond the curvature
singularity at $r=0$ have also been put forward by other authors
\cite{syng,lyka}. In fact there is a small and -- in our opinion undeservedly
-- little-known tradition in this field; we will therefore also rederive with
new tools and in a slightly different perspective some results that are in
principle not quite new, but not so well known. Given the renewed interest in
extensions of back hole solutions motivated by string theory and taking into
account ``the resistance to any change in the rules" \cite{lyka}
we present them in this letter as well.

Compared to previous work, our treatment is simpler both from a
conceptual and from a computational point of view. While in \cite{lyka} similar
results have been obtained by embedding Schwarzschild \st\ in a cosmical dust
solution, our investigations rely on the introduction of several \cod\
systems for the Schwarzschild solution which to our knowledge have not been
presented so far. As simple \cod\ systems have frequently triggered
progress in general relativity, this might be an independent point of interest
of our results. We will
state carefully the rules for the extension of the geodesics;
in particular we will point out a subtle difference between the \twod\
reduction of the geometry obtained by suppressing the angular part
and the full \fourd\ geometry. This difference implies that in contrast to the
\twod\ reduction where the extension is still a topological manifold -- albeit
one for which the metric structure is divergent at infinitely many isolated
\oned\ submanifolds -- it is necessary in the full \fourd\
geometry to replace the structure of a manifold by the closely related
structure
of a {\em stratified variety}.

In the rest of this letter we first introduce new local
\cod s for Schwarzschild \st. In these \cod s the geometry near the
singularity is explored first by use of radial geodesics; in a second step
also non-radial geodesics are considered. We conclude with a description of
the global features of this extension of the Schwarzschild solution.
\vskip 5mm

The Schwarzschild metric is the locally unique solution of the Einstein
equations in empty space \cite{KS,Dr} describing a spherically symmetric
gravitational field, static outside the horizon $r_H = 2M$. In terms of the
original $(r,t)$ \cod s the solution is singular at the horizon, but
other \cod\ systems are known extending the solution smoothly across the
horizon all the way up to the curvature singularity at $r = 0$. The most
complete description of this kind is the one by Kruskal \cite{Kr} and
Szekeres \cite{Sz}. This \st\ is {\em not} geodesically complete; however
it has the important property that the only obstruction to geodesic
completeness are the curvature singularities: all geodesics
can be extended indefinitely unless they reach the curvature singularity.

The line element characterising the \ksz\ solution takes the form

\begin{equation}
\frac{\rd s^2}{4M^2}\, =\, f(u,v)\, \left( - \rd v^2 + \rd u^2 \right)\, +\,
g(u,v)\, \rd\Omega^2,
\labl{1}

\noindent
where $f(u,v)$ and $g(u,v)$ are functions of $u^2 - v^2$ only, and the
connection with the standard Schwarzschild \cod\ is

\begin{equation}
u^2 - v^2 = \left( \frac{r}{2M} - 1 \right)\, \eE^{r/2M}, \hspace{3em}
\frac{v}{u}\, =\, \left\{ \begin{array}{ll}
                          \tanh \frac{t}{4M}, & \mbox{if } |v| < |u|; \\
                            &  \\
                          \coth \frac{t}{4M}, & \mbox{if } |v| > |u|.
                          \end{array} \right.
\labl{2}

\noindent
The functions $f$ and $g$ are then implicitly defined by

\begin{equation}
f(u,v) = \frac{8M}{r}\, \eE^{-r/2M}, \hspace{3em} g(u,v) = \frac{r^2}{4 M^2}.
\labl{3}

\noindent
The \st\ domain covered by this solution of the Einstein equations
consists of two physically distinct exterior regions, extending from $r =
\infty$ to the horizon and usually denoted as regions I and III, and two
interior regions bounded by the horizon and either the future or the past
singularity, known as regions II and IV, respectively \cite{Ch}.

A new \cod\ system for the Schwarzschild solution is given for the two
regions I, III outside the \bh\ horizon $(r > 2M)$ by the line element

\begin{equation}
\rd s^2\, =\, -\tanh^2 \frac{\rho}{2}\, \rd t^2\, +\, 4 M^2 \cosh^4
           \frac{\rho}{2}\, \left( \rd \rho^2 +
           \rd \Omega^2 \right).
\labl{18}

\noindent
It is easily checked that this is a solution of the matter-free Einstein
equations with $r > 2M$ by performing the transformation to standard $(r,t)$
\cod s:

\begin{equation}
\frac{r}{2M}\, =\, \cosh^2 \frac \rho 2.
\labl{19}

\noindent
The explicit connection of $\rho$ with the \ksz\ \cod s is

\begin{equation}
u = \eE^{1/2 \cosh^2 \frac{\rho}{2} } \sinh
    \frac{\rho}{2} \cosh \frac{t}{4M},            \hspace{3em}
v = \eE^{1/2 \cosh^2 \frac{\rho}{2} } \sinh
    \frac{\rho}{2} \sinh \frac{t}{4M}.
\labl{20}

\noindent
In the domain $\rho=\rho_+ > 0$ this corresponds to $u > 0$, and one obtains a
cover of region I in the \ksz\ diagram. Setting $\rho = \rho_- <0$,
$\rho_-$ covers the exterior region III, with
$u < 0$. The double covering of the exterior region $r > 2M$ in the
\ksz\ diagram is well-known to describe two physically distinct
sheets of Schwarzschild \st.

We observe that the \cod\ system (\ref{18}) can be continued
to the interior region of Schwarzschild \st\ by a simple Wick rotation.
\begin{equation}
\rho\, =  \mp \:\ii \: \rho_{\pm}.
\labl{17}

\noindent
The two possible choices for the sign in the Wick rotation correspond
to two possible ways of patching regions outside the horizon with $r>2M$ to
the interior of a \bh, yielding region I and III respectively.
The line element inside the black-hole horizon $(r < 2M)$ is then given by
\begin{equation}
\rd s^2\, =\, \tan^2 \frac{\rho}{2}\, \rd t^2\, +\, 4 M^2 \cos^4
\frac{\rho}{2}\, \left( - \rd \rho^2 + \rd \Omega^2 \right),
\labl{4}

\noindent
for $0 \leq \rho \leq 2 \pi$. This shows that after the Wick rotation $\rho$
becomes a time-like and $t$ a space-like \cod, analogously to ordinary
Schwarzschild \cod s.  Comparing with equation (\ref{19}), the connection
with the usual $(r,t)$ \cod s is made by the transformation

\begin{equation}
\cos^2 \frac{\rho}{2}\, =\, \frac{r}{2M}.
\labl{5}

\noindent
Clearly, this provides a double-valued parametrization of the interior region
containing the singularity $r = 0$, or $\rho = \pi$: every value $r > 0$
corresponds to two distinct values of $\rho$ in the domain $\left[ 0, 2 \pi
\right]$. This double-valuedness holds both near the past and the future
horizon, as follows from the transformation to \ksz\ \cod s

\begin{equation}
u = \eE^{1/2 \cos^2 \frac{\rho}{2}}\, \sin \frac{\rho}{2}\,
    \sinh \frac{t}{4M}, \hspace{3em}
v = \eE^{1/2 \cos^2 \frac{\rho}{2}}\, \sin \frac{\rho}{2}\,
    \cosh \frac{t}{4M}.
\labl{6}

\noindent
In order to interpret the double covering of the region inside the horizon
by our solution, we are going to investigate time-like geodesics. For
simplicity let us start with incoming radial geodesics; they are
solutions of the equation

\begin{equation}
\frac{\rd\rho}{\rd t}\, =\, \frac{- \sin \frac{\rho}{2}}{2M \cos^4
\frac{\rho}{2}},
\labl{7}

\noindent
from which we deduce the following expression for the proper time measured by
an infalling test particle

\begin{equation}
\rd\tau\, =\, 2M\, \cos^2 \frac{\rho}{2}\, \sin \frac{\rho}{2}\, \rd\rho,
\labl{8}

\noindent
with $\tau$ increasing as $\rho$ increases from 0 at the horizon to $\pi$
at the singularity.

Moving now on to values $\rho > \pi$ and increasing $\tau$ according to
eq.(\ref{8}) we see that the particle moves away from the singularity as its
proper time increases, and the region of $(u,v)$ space we are in contains a
{\em past} singularity; hence it is physically distinct from the region before
the encounter with the infinite curvature singularity. The double-valuedness
thus connects the interior of the original Schwarzschild \st\ at the curvature
singularity with a new region that can be interpreted as the inner region
of the white hole of a new Schwarzschild \st. We will discuss the global
features of such an extended \st\ in more detail below.

We emphasize that we are {\em not} extending the $r$ \cod\ to
negative values, nor the \ksz\ \cod s to $v^2 - u^2 > 1$.
Rather, we patch a new, physically distinct, region of positive $r$-values
to \st, which can be reached only by passing the curvature singularity.

In the sequel, we show that this extension can be done without
running into physical paradoxes.
It follows from eq.(\ref{8}) that two points
on a geodesic separated by the singularity are only a finite interval of proper
time apart \cite{syng}:

\begin{equation}
\int_{\rho = \pi - \varepsilon}^{\rho = \pi + \varepsilon} \rd\tau\, =\,
 \frac{8M}{3} \cos^3 \left( \frac{\pi - \varepsilon}{2} \right)\, \approx\,
 \frac{M}{3}\, \varepsilon^3.
\labl{9}

\noindent
We also observe that the geodesic is smooth everywhere in the sense that the
velocity as measured in the \ksz\ \cod s is finite and
continuous, even when the proper velocity and acceleration become momentarily
infinite, and that the light-cone structure is preserved upon crossing the
singularity.

In support of these statements, we consider a system of test particles
falling in radially from rest at any point $0 < r_0 < \infty$. To construct
the corresponding geodesics it is most convenient to introduce a collection
of \cod\ systems $(\kappa, t, \Omega)$ parametrized by a variable $\kappa_H$
of the form

\begin{equation}
\rd s^2\, =\, -\, \frac{\kappa^2 - \kappa_H^2}{(1 + \kappa_H^2) \kappa^2}\,
\rd t^2\,
  +\, 4 M^2\, \frac{(1 + \kappa_H^2)^2 \kappa^4}{\kappa_H^4 (1 + \kappa^2)^2}\,
  \left[ \frac{1 + \kappa_H^2}{(1 + \kappa^2)^2} \frac{4 \rd\kappa^2}{\kappa^2
-
         \kappa_H^2} + \rd \Omega^2 \right].
\labl{10}

\noindent
These \cod\ systems describe Schwarzschild \st\ for $r < r_0$
via the transformation

\begin{equation}
\frac{r}{2M}\, =\, \frac{r_0}{2M}\, \frac{\kappa^2}{1 + \kappa^2},
\labl{11}

\noindent
with $r_0$ related to the parameter $\kappa_H$, the value of $\kappa$ at the
horizon, by

\begin{equation}
\frac{r_0}{2M}\, =\, \frac{1 + \kappa_H^2}{\kappa_H^2}.
\labl{12}

\noindent
Just like the $(\rho, t)$ \cod\ system, the $(\kappa, t)$ \cod\
systems represent a double cover of the domain $0 < r < r_0$, and in fact
extend into the new region of \st\ beyond the singularity. In terms of
$\kappa$ the equation for radial geodesics becomes

\begin{equation}
\frac{1}{4M}\, \frac{\rd t}{\rd\kappa}\, =\, \frac{(1 + \kappa_H^2)^2}
{\kappa_H^3}\, \frac{\kappa^4}{(1 + \kappa^2)^2 (\kappa_H^2 - \kappa^2)}.
\labl{13}

\noindent
For $\kappa < \kappa_H$ this has the solution

\begin{equation}
\frac{t - t_0}{4M}\, =\, \mbox{arctanh} \frac{\kappa}{\kappa_H}\, -\,
\left(\frac{1 + 3 \kappa_H^2}{2\kappa_H^3} \right)\, \arctan \kappa\, +\,
\left(\frac{1 + \kappa_H^2}{2\kappa_H^3}\right)\, \frac{\kappa}{1 + \kappa^2}.
\labl{14}

\noindent
It follows that the horizon at $t \rightarrow \infty$ is crossed at $\kappa =
\kappa_H$, whilst the geodesic reaches the singularity $\kappa = 0$ at $t =
t_0$. Inside the horizon the velocity of the test particle in the
\ksz\ \cod\ system is found to be

\begin{equation}
\frac{\rd u}{\rd v}\, =\, \tanh \left[ \frac{t_0}{4M} - \left(\frac{1 + 3
\kappa_H^2}{
      2 \kappa_H^3} \right) \arctan \kappa + \left( \frac{1 + \kappa_H^2}{
      2 \kappa_H^3}\right) \frac{\kappa}{1 + \kappa^2} \right].
\labl{15}

\noindent
Therefore $|\rd u/ \rd v| < 1$ at all times, both for $\kappa > 0$ (before
reaching the singularity) and for $\kappa < 0$ (after traversing the
singularity). On the other hand, null geodesics (light rays) correspond to
straight lines

\begin{equation}
\frac{\rd u}{\rd v}\, =\, \pm 1.
\labl{16}

\noindent
It follows that time-like geodesics remain time-like also in the region of
space-time corresponding to $\kappa < 0$. Equation (\ref{15}) proves the above
assertion regarding the finiteness and continuity of the velocity in the
$(u,v)$ \cod\ system for all values of $\kappa$.

So far we have only considered radial geodesics; our results show that the
\twod\ geometry obtained by suppressing the angular degrees of freedom
looks as follows: we obtain an infinite sequence of (\twod\ reductions of)
\ksz\ \st s as depicted in figure \ref{kch} which forms a smooth topological
manifold. However, the metric structure of this manifolds diverges along
infinitely many \oned\ submanifolds, marked by bold lines in figure \ref{kch}.
It can be shown that also the black hole solution
for \twod\ dilaton gravity can be extended similarly \cite{peet}.
\begin{figure}[tbh]
\epsfysize=10cm
\begin{center}
\leavevmode
\epsfbox{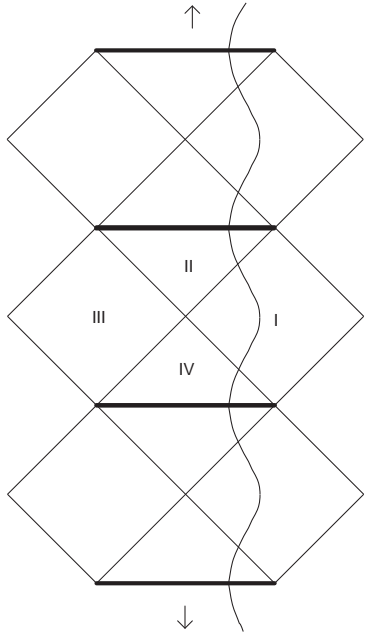}
\caption{Schematic representation of the global structure of the extension
of Schwarzschild \st: an infinite sequence of \ksz\ domains which are connected
along lines of infinite curvature. The line depicts a radial geodesic
crossing the singularities as described in the text.}
\label{kch}
\end{center}
\end{figure}

In a second step we are going to explore the extension of the angular
\cod s $\varphi$ and $\theta$; let us therefore now have a look at
non-radial geodesics. It is crucial to remark that angular momentum is a
conserved quantity in our problem. Using this, we can choose
$\theta = \frac \pi 2 = {\rm const}$ on both sheets. Chosing this constant to
be the same amounts to imposing the conservation law for angular momentum
across the singularity.

In addition, we have the conserved quantity

\begin{equation}
L=\cos^4\frac{\rho}{2}\frac{\rd\varphi}{\rd\tau} .
\end{equation}

\noindent
which allows to derive for geodesics with non-vanishing angular momentum a
relation between $\rho$ and $\varphi$ that can be integrated across the
singularity.

For geodesics in the equatorial plane entering from infinity we obtain

\begin{equation}
\left( \frac{\rd\rho/\rd\tau}{\rd\varphi/ \rd\tau} \right)^2 =
1+\frac{4M^2 \cos^4\frac{\rho}{2}}{L^2\sin^2\frac{\rho}{2}} ,
\end{equation}

\noindent
which is finite in a neighbourhood of the singularity and becomes
$1$ in the limit $\rho\rightarrow\pi$. This implies
the existence of a non-singular relation $\rho=\rho(\varphi)$, so
that $\varphi$ can be integrated to a smooth function of proper time.
For geodesics entering from a finite distance, an analogous argument goes
through as well. However, the equations become more complicated and we
refrain from presenting them in detail.

The singularity can thus be  interpreted as a single point on time-like
geodesics at which the proper acceleration becomes momentarily infinite, as in
the case of a test charge moving through the Coulomb singularity of a fixed
charge of opposite sign. Just as in the electric analogue, a conservation law,
the conservation of energy-momentum on geodesics, guarantees that in the
absence of a hard (delta-function) core in the potential the particle continues
its worldline after moving through the singular point, with finite velocity.

{}From this analogy, one might speculate that some of these features might
survive in a quantum theory where it is well-known that
singularities like the Coulomb singularity do not exclude the existence of
well-behaved, normalizable particle wave-functions.
Further support for this conjecture comes from recent work on \twod\
black-hole models emerging from string theory, which are described by
gauged Wess-Zumino-Witten sigma models \cite{W,DV2}.
These theories also exhibit a pattern of geodesic continuation of \st\ to
new regions. Our results show that already in purely classical gravity such
\st\ extensions occur, even in the presence of curvature singularities.

Observing that in the new region of \st\ there is a past singularity,
it follows that eventually the geodesics reach another event-horizon, where
the solution (\ref{4}) can again be continued to an exterior region with a
metric structure of type (\ref{18}), but this time the Wick rotation is made
around the point $2\pi$, rather than $\rho = 0$. Patching together these
solutions in a new \ksz\ \cod\ system, one encounters a new
future singularity, and the construction described here can be repeated
indefinitely, both in the forward and backward directions of proper time. This
is particularly clear if one follows one of the geodesics starting from rest at
a finite distance $r_0$, as described by eq.(\ref{14}); passing through the
singularity at $r = 0$ it reaches again a maximal distance $r_0$ in the new
sheet of \st, then falls back in until it reaches the next singularity,
etc. In order to avoid problems with causality due to closed time-like loops we
cannot identify the new regions with any earlier ones, a procedure which would
result in a covering of \st\ by a finite number of \ksz\ \st s.
We are thus forced to consider an infinite covering.

We are now in a position to describe the global geometric features of this
extension of Schwarzschild \st: it consists of an infinite sequence of
\ksz\ domains, connected along lines of infinite curvature. This geometrical
object is {\em not} a manifold any more, but rather a {\em stratified variety};
by definition, a stratified variety is a connected topological space which
can be represented as the disjoint sum of manifolds which can have different
dimensions. The manifolds that make up this variety are referred to as {\em
strata}. A simple example of a stratified variety is a double cone:
it has a \twod\ stratum consisting of two parts that are isomorphic separately
to an infinitely long cylinder, and a \zerod\ stratum, the tip of the cone.

In our case there are two strata: a \fourd\ one consisting of a disjoint sum
of countably many \ksz\ \st s, and \oned\  stratum  which consists of the
singular lines (marked by the bold lines in figure \ref{kch}).
In fact, there is a \twod\ sphere of radius $r$ attached to any point
of the Kruskal diagram which degenerates at the singularity $r=0$, the
\oned\ stratum.
The strata are glued together such that  \st\ as a whole is connected; as we
have seen, any timelike geodesic can be continued naturally through the
singular stratum. This again has an analogue for the cone
(supplied with the flat metric on it): there are two types of geodesics on
a cone, curves that could be described as slightly curved hyperbolae and
straight lines through the tip. The latter can be extended smoothly to the
second sheet of the cone. However, in contrast to Schwarzschild \st, only
these radial geodesics cross the singularity.

In the stratified variety we propose as the extension of Schwarzschild \st\
{\em all} geodesics can be continued indefinitely in a unique way, including
those that reach the curvature singularity; therefore the resulting \st\ is
geodesically complete and should be seen as the truly inextendable \st\
underlying Schwarzschild geometry.

\end{document}